\documentclass[aps,preprint,showpacs,superscriptaddress]{revtex4}
\usepackage{amssymb}
\usepackage{graphicx}
\usepackage{dcolumn}
\usepackage{bm}

\begin{document}

\title{Electron-positron pair production in an arbitrary polarized ultrastrong laser field}
\author{Liang-Yong He}
\affiliation{Key Laboratory of Beam Technology and Materials
Modification of the Ministry of Education, Beijing Normal
University, Beijing 100875, China}
\affiliation{College of Nuclear Science and Technology, Beijing
Normal University, Beijing 100875, China}
\author{Bai-Song Xie\footnote{Corresponding author. Email address: bsxie@bnu.edu.cn}}
\affiliation{Key Laboratory of Beam Technology and Materials
Modification of the Ministry of Education, Beijing Normal
University, Beijing 100875, China}
\affiliation{College of Nuclear Science and Technology, Beijing
Normal University, Beijing 100875, China}
\author{Xin-Heng Guo}
\affiliation{Key Laboratory of Beam Technology and Materials
Modification of the Ministry of Education, Beijing Normal
University, Beijing 100875, China}
\affiliation{College of Nuclear Science and Technology, Beijing
Normal University, Beijing 100875, China}
\author{Hong-Yu Wang}
\affiliation{Department of Physics, Anshan Normal University,
Anshan 114005, China}

\begin{abstract}
Electron-positron pair production in an arbitrary polarized
ultrastrong laser field is investigated in the first order
perturbation approximation in which the Volkov states are used for
convenient calculation of scattering amplitude and cross section.
It is found surprisingly that the optimal pair production depends
strongly on the polarization. For some cases of field parameters,
the optimal field is elliptically polarized or evenly circularly
polarized one, rather than the usual linear polarization as
indicated by previous works. Some insights into pair generation
are given and some interesting unexpected features are also
discussed briefly.
\end{abstract}

\pacs{12.20.Ds, 13.40.-f, 32.80.Wr, 42.50.Ct}
\date{\today}
\maketitle

\section{Introduction}

Strong field physics has been an active research field since the
invention of the technique chirped pulse amplification (CPA) in
1985 \cite{StMo}, which draws the intensity of tabletop laser
from gigwatt to erawatt \cite{MoTaBu} and makes it possible to
test the quantum electrodynamics (QED) theory in laboratories
\cite{Burke}. The Stanford Linac Acceleration (SLAC) experiment
in 1997 \cite{Burke, Bamber} have observed electrons-positrons ($e^+ e^-$) pair
production and revealed nonlinear QED effects, which have
resurrected research interests in pair creation in strong fields.
The forthcoming extreme-light-infrastructure (ELI)
\cite{ELI} will surely boost breakthrough in strong field physics
research. This will be the strongest field in the world, though
its intensity is approximately about $10^{25} \rm{W/cm^2}$ and still 4
orders less than the Schwinger's critical field intensity $10^{29} \rm{W/cm^2}$
\cite{Schwinger}.

Several methods in both of theory and numeric calculation have
been proposed to analyze pair production in external fields since
1930s \cite{Sauter,
Volkov,KimPa,DuSc,GiKl,DuQHWGi,MVG,DeM,MHaKe,HuMKe,Bulanov}.
One of the most important approaches is the perturbation method in
which the corresponding Volkov states are used as bases. It was
pioneered by Volkov \cite{Volkov}, and developed by Reiss
\cite{Reiss}, Nikishov \cite{NiRi},
M\"{u}ller \cite{MVG,DeM,MHaKe,HuMKe}, Keitel
\cite{MHaKe,HuMKe} and so on. But all of the above studies
are based on particular polarized laser fields with either
circular or linear polarization
\cite{Reiss,NiRi,MVG,DeM,Bulanov,MHaKe,HuMKe}. Most studies
have manifested that the pair production rates are larger in
linearly polarized fields compared than that in
circular ones \cite{Altarelli,DeM,Popov,Bulanov} in some conditions.
Albeit these progresses which have been achieved, no complete analytical or numerical
studies have been performed to reveal the influence of laser
fields with arbitrary polarization on pair creation, especially in
the framework of perturbation method even though this method seems
simple and valid for the problem. The reason may be that
relatively tedious derivation of scattering amplitude and cross
section is involved in that case of elliptic polarization where
the scattering partial waves have many coupled terms.

Therefore, it is worthwhile to fill this research gap by exploring
the relation between the elliptically polarized laser fields and
pair production, which will give more insight into the SLAC
experiment. Our study will show
that a better pair production by elliptically polarized laser
field can occur possibly. If we have got more information and
recognition, it will be helpful not only to uncover more phenomena
but also to regulate involved parameters in future experiments.

In the present paper, we will deal with one of these problems,
through difficult theoretical and computing calculations. The case
of head-on collision, occurring between a high-energy gamma photon
and multiple photons of the laser fields, is considered. This
process is called multi-photon reaction. Fortunately, we get the
strict result in the framework of first order perturbation theory,
and give thoroughly numerical calculations of some fixed
parameters. There is, in fact, a surprising result that the pair
creation relies strongly on the parameters of the laser fields,
especially $\Upsilon$ (describing the elliptical polarization) in
the condition that $\eta_0$ (the Lorentz-invariant dimensionless
strength parameter \cite{GrRe}) could be compared with
$\vartheta$ (the collision energy). The optimal filed is not the
linearly polarized field, but in some cases it is elliptically
polarized and sometimes even the circularly polarized.

The paper is organized as follows. Firstly, in Sec.II, we give a
concrete theoretical derivation of the total cross section in the
first order perturbation theory using the Volkov states wave
functions. In Sec. III, we present some demonstrations of
numerical results which show clearly and vividly how pair creation
is influenced by a few important parameters such as the
polarization $\Upsilon$, the Lorentz-invariant normalized field
strength $ \eta_0$, and the collision energy $\vartheta$. The main
results will be summarized and some insights and discussions are
given briefly in the final section.

\section{Theoretical formalism}

In this section, we give a detailed derivation of the total pair
production in the interaction between strong laser field and high
energy photon. Some techniques used here are employed and
developed from Ref. \cite{GrRe} and Ref. \cite{BeLiPi}.
First of all, some notations must be employed for simplicity and
convenience. Throughout the paper, an elliptically polarized laser
field, $A^{\mu}=a[{\varepsilon}^{\mu}_1 \cos{(k\cdot x)}+\Upsilon
{\varepsilon}^{\mu}_2 \sin{(k\cdot x)}]$, is adopted, where $a$ is
the amplitude of vector potential $A$ of laser field, $k^2=0$,
$k^{\mu}={\omega}(1,0,0,1)$, ${\varepsilon}^{\mu}_{1}=(0,1,0,0)$,
${\varepsilon}^{\mu}_{2}=(0,0,1,0)$ and $-1 \le \Upsilon \le 1$.
The natural units, $\hbar = c =1$, are also used.

We define $\Lambda = \frac{1}{2}(1+{\Upsilon}^2)$, the average number density of photons
${\rho}(\omega) = \left\langle \frac{a^2
\omega}{4\pi}[1+({\Upsilon}^2-1)\cos^2{(k \cdot x)}] \right\rangle
= \frac{a^2 \Lambda \omega}{4\pi}$ and $\eta = \frac{e}{m_0}
\sqrt{\left| \left\langle A_{\mu}A^{\mu} \right\rangle \right|} =
\frac{ea}{m_0} \sqrt{\Lambda}$, where $e$ and $m_0$ are
respectively the charge and the rest mass of an electron. $\eta_0$
is defined as $\frac{ea}{m_0}$. The scalar four-product is defined
as $A \cdot B = A_{\mu}B^{\mu} = A_0 B^0 - A_i B^i$ and
$/\kern-0.70em A = A_{\mu} {\gamma}^{\mu}$, in which
${\gamma}^{\mu}$ is the standard Dirac matrices. We also sometimes
use $\phi$ in place of $k \cdot x$ for brevity.

The solution of electrons/positrons under this type field could be
well solved in terms of the known Volkov states \cite{Volkov,
GrRe}, i.e. for electron it is $ \Psi_{p,s}^{(e)} = N_p \left[
1+ \frac{e{/\kern-0.40em k}{/\kern-0.40em A}}{2(p\cdot k)} \right]
u(p,s)\exp[iS^{(e)}_{(p,s)}(k\cdot x)] $ and for positron it is
$\Psi_{\bar{p},-\bar{s}}^{(e^+)} = N_{\bar{p}} \left[ 1-
\frac{e{/\kern-0.40em k}{/\kern-0.40em A}}{2({\bar{p}}\cdot k)}
\right]
v(\bar{p},-\bar{s})\exp[iS^{(e^+)}_{(\bar{p},-\bar{s})}(k\cdot
x)]$. Here $u(p,s)$ and $v(\bar{p},-\bar{s})$ are respectively the
unit spinor of a free electron and a free positron, $p$ and
$\bar{p}$ the momenta, $s$ and $\bar{s}$ the spins, and
$S^{(e)}_{(p,s)}$ and $S^{(e^+)}_{(\bar{p},-\bar{s})}$ are
\begin{equation}
            \label{eq1}
            \left\{
            \begin{array}{l}
                S^{(e)}_{(p,s)} = -p\cdot x - \int\limits_0^{k\cdot x} \left[
                \frac{e(p\cdot A)}{p\cdot k} - \frac{e^2 A^2}{2(p\cdot k)}
                \right]\,d\phi, \ \\
                S^{(e^+)}_{(\bar{p},-\bar{s})} = \bar{p}\cdot x - \int\limits_0^{k\cdot x} \left[
                \frac{e(\bar{p}\cdot A)}{\bar{p}\cdot k} + \frac{e^2 A^2}{2(\bar{p}\cdot k)}
                \right]\,d\phi. \
            \end{array}
            \right.
\end{equation}
We will denote $S^{(e^+)}_{(\bar{p},-\bar{s})}$ simply as
$S^{(e^+)}$ and $S^{(e)}_{(p,s)}$ as $S^{(e)}$ in the following
discussion. Then, the amplitude for pair production process due to
the laser field collision with a high energetic gamma photon of
$A^{'}=N_0\varepsilon^{'}\exp{(-ik^{'}\cdot x)}$ is the following
in the framework of first order perturbation theory:
        $$
        \begin{array}{rcl}
        S_{fi} & = & -ie\int\,d^4x\ \bar{\Psi}_{p_f,s_f}^{(e)}(x) {/\kern-0.75em A}^{'}(x) \Psi_{\bar{p}_i,-\bar{s}_i}^{(e^+)} (x) \\
               & = & -ieN_{p_f}N_{\bar{p}_i}N_0 \int\, d^4x \bar{u}(p_f, s_f)Mv(\bar{p}_i, -\bar{s}_i)\exp\left[ iS^{(e^+)}(x) - iS^{(e)}(x) -ik^{'} \cdot x \right].
        \end{array}
        $$
Through a detailed calculation, we have
        $$
        \begin{array}{rcl}
        S^{(e^+)}(x) - S^{(e)}(x) - k^{'}\cdot x & = & \left[ q_{f;\mu} + \bar{q}_{i; \mu} - k^{'}_{\mu}  \right] x^{\mu} +ea \left[ \frac{q_{f}\cdot \varepsilon_1}{q_f\cdot k} - \frac{\bar{q}_i\cdot \varepsilon_1}{\bar{q}_i\cdot k}  \right]\sin{\phi} \\
               &   & -\Upsilon ea \left[ \frac{q_{f}\cdot \varepsilon_2}{q_f\cdot k} - \frac{\bar{q}_i\cdot \varepsilon_2}{\bar{q}_i\cdot k}  \right] \cos{\phi} - \frac{\Lambda - 1}{4}e^2a^2\sin{(2\phi)} \left[ \frac{1}{q_f\cdot k}+\frac{1}{\bar{q_i}\cdot k}  \right] \\
               & = & \left[ q_{f;\mu} + \bar{q}_{i; \mu} - k^{'}_{\mu}  \right] x^{\mu} + eaQ\cdot\varepsilon_1\sin{\phi} - \Upsilon ea Q\cdot\varepsilon_2 \cos{\phi} \\
               &   & - \frac{\Lambda - 1}{4}Ge^2a^2\sin{(2\phi)} \\
               & = &  \left[ q_{f;\mu} + \bar{q}_{i; \mu} - k^{'}_{\mu}  \right] x^{\mu} - z\sin{(\phi - \phi_0)} -y\sin{(2\phi)}.
        \end{array}
        $$
Here, we have defined the following quantities:
\begin{equation}
            \label{eq2}
            \left\{
            \begin{array}{l}
                 q^{\mu} = p^{\mu} + \Lambda \frac{e^2 a^2}{2p \cdot k} k^{\mu},
                 Q = \frac{q_f}{q_f \cdot k}  - \frac{ {\bar{q}}_i }{ {\bar{q}}_i \cdot k }, \\
                 ea(Q\cdot {\varepsilon}_1) \sin{\phi} - \Upsilon ea (Q\cdot {\varepsilon}_2) \cos{\phi} = -z \sin{(\phi - {\phi}_0)}, \\
                 z = ea \sqrt{{\left( Q\cdot {\varepsilon}_1 \right)}^2 + {\left( \Upsilon Q\cdot {\varepsilon}_2 \right)}^2}, \\
                 Q\cdot {\varepsilon}_1 = -\frac{z\cos{{\phi}_0}}{ea}, \Upsilon Q\cdot \varepsilon_2 = -\frac{z\sin{{\phi}_0}}{ea}, \\
                 G = \frac{1}{q_f \cdot k} + \frac{1}{ {\bar{q}}_i \cdot k }, \\
                 y = \frac{\Lambda-1}{4}Ge^2a^2.
            \end{array}
            \right.
\end{equation}
Thus, we find the amplitude
        $$
        \begin{array}{rcl}
        S_{fi} & = & -ieN_{p_f}N_{\bar{p}_i}N_0\int d^4x \ \bar{u}(p_f, s_f)Mv(\bar{p}_i, -{\bar{s}_i}) \times \\
               &   & \exp\left\{ i\left[ q_{f;\mu} + \bar{q}_{i; \mu} - k^{'}_{\mu}  \right] x^{\mu} - iz\sin{(\phi - \phi_0)} -iy\sin{(2\phi)}  \right\} \\
               & = & -ieN_{p_f}N_{\bar{p}_i}N_0 \sum\limits_{m=-\infty}^{\infty} J_m(y) \sum\limits_{n=-\infty}^{\infty}(2\pi)^4 \delta\left( q_{f;\mu} + \bar{q}_{i; \mu} - k^{'}_{\mu} -2mk_{\mu} -nk_{\mu} \right) \times \\
               &   & \bar{u}(p_f, s_f) \left[ B_n(z)M_0 + B_{1n}(z)M_1 + B_{2n}(z)M_2 + B_{3n}(z)M_3  \right] v(\bar{p}_i, -\bar{s}_i),
        \end{array}
        $$
where some notations have been adopted as
\begin{equation}
            \label{eq4}
            \left\{
            \begin{array}{l}
                M = M_0 + M_1 \cos{\phi} + M_2 \sin{\phi} + M_3 \cos{ (2\phi)}, \\
                M_0 = { /\kern-0.40em {\varepsilon} }^{'} - \Lambda e^2 a^2 \frac{ (k \cdot {\varepsilon}^{'}) {/\kern-0.40em k}}{2 (p_f \cdot k) ( {\bar{p}}_i \cdot k)}, \\
                M_1 = ea \left[\frac{{ {/\kern-0.30em \varepsilon}_1 } {/\kern-0.30em k} { {/\kern-0.30em \varepsilon}^{'} }}{2(p_f \cdot k)}
                - \frac{ {/\kern-0.30em \varepsilon}^{'} { /\kern-0.30em k} {/\kern-0.30em \varepsilon}_{1} }{2({\bar{p}}_i \cdot k)} \right], \\
                M_2 = \Upsilon ea \left[\frac{{ {/\kern-0.30em \varepsilon}_2 } {/\kern-0.30em k} { {/\kern-0.30em \varepsilon}^{'} }}{2(p_f \cdot k)}
                - \frac{ {/\kern-0.30em \varepsilon}^{'} { /\kern-0.30em k} {/\kern-0.30em \varepsilon}_{2} }{2({\bar{p}}_i \cdot k)} \right], \\
                M_3 = (\Lambda - 1) e^2 a^2 \frac{ (k \cdot {\varepsilon}^{'}) {/\kern-0.30em k}}{2 (p_f \cdot k) ( {\bar{p}}_i \cdot k)},
            \end{array}
            \right.
\end{equation}
and some special functions like $J_{n}(y)$ and $B_n(z)$ etc are
given by
\begin{equation}
            \label{eq5}
            \left\{
            \begin{array}{l}
                \exp{\left[ -iy\sin{(2\phi)}  \right]} = \sum\limits_{n=-\infty}^{\infty} J_n(y)\exp{\left[ -i2n\phi  \right]}, \\
                \exp{\left[ -iz\sin{(\phi - \phi_0))}  \right]} = \sum\limits_{n=-\infty}^{\infty} B_n(z)\exp{\left[ -in\phi  \right]}, \\
                \cos{\phi} \exp{\left[ -iz\sin{(\phi - \phi_0)}  \right]} = \sum\limits_{n=-\infty}^{\infty} B_{1n}(z)\exp{\left[ -in\phi  \right]}, \\
                \sin{\phi} \exp{\left[ -iz\sin{(\phi - \phi_0)}  \right]} = \sum\limits_{n=-\infty}^{\infty} B_{2n}(z)\exp{\left[ -in\phi  \right]}, \\
                \cos{(2\phi)} \exp{\left[ -iz\sin{(\phi - \phi_0)}  \right]} = \sum\limits_{n=-\infty}^{\infty} B_{3n}(z)\exp{\left[ -in\phi
                \right]}.
            \end{array}
            \right.
\end{equation}
The amplitude can be also rewritten in a more compact form,
\begin{equation}
            \label{eq3}
        \begin{array}{rcl}
        S_{fi} & = & -ieN_{p_f}N_{\bar{p}_i}N_0 \times \\
               &   & \sum\limits_{m=-\infty}^{\infty} J_m(y) \sum\limits_{n=-\infty}^{\infty}(2\pi)^4 \delta\left( q_{f;\mu} + \bar{q}_{i; \mu} - k^{'}_{\mu} -2mk_{\mu} -nk_{\mu} \right) \bar{u}(p_f, s_f) M_n v(\bar{p}_i, -\bar{s}_i),
        \end{array}
\end{equation}
where
$$
M_n = B_n(z)M_0 + B_{1n}(z)M_1 + B_{2n}(z)M_2 + B_{3n}(z)M_3.
$$

What we are interested in is $\overline{{\left\vert S_{fi}
\right\vert}^2}$. With a deliberate consideration on Eq.
(\ref{eq3}), the main task has been converted to how to tackle
$M_{fi,n} = \bar{u}(p_f, s_f)M_nv(\bar{p}_i, -\bar{s}_i)$. Then we
will have to deal with $M_{fi,n}M_{fi,m}^{*}$ in the general
elliptic polarized field rather than $M_{fi,n}M_{fi,n}^{*}$ in the
circular polarized field \cite{GrRe} because there are two
series summations existed in Eq. (\ref{eq3}) in the case of
general elliptic polarization.

The final total cross section $\bar{\sigma}$ is obtained by
averaging over the photon polarization $\lambda$ and summing over the
$e^+ e^-$ spins, $\overline{M_{fi,n}M_{fi,m}^{*}} = \frac{1}{2}\sum\limits_{\lambda} \sum\limits_{\bar{s}_i, s_f}M_{fi,n}M_{fi,m}^{*}$,
\begin{equation}
        \label{eq7}
        \begin{array}{rcl}
          \overline{M_{fi,n}M_{fi,m}^{*}} & = & \frac{1}{2}\sum\limits_{\lambda} Tr \left[ \frac{{/\kern -0.50em p}_f + m_0}{2m_0}M_n \frac{{/\kern -0.50em {\bar{p}}}_i - m_0}{2m_0} \gamma^0 M_m^{\dagger} \gamma_0 \right] \\
          & = & \frac{1}{8m_0^2} \sum\limits_{\lambda} Tr\left[ (/\kern-0.40em p_f + m_0)M_n(/\kern-0.40em \bar{p}_i - m_0) \gamma^0 M_m^{\dagger} \gamma^0 \right].
        \end{array}
\end{equation}
Before proceeding, we define some notations as below
\begin{equation}
        \label{eq8}
        \left\{
        \begin{array}{rcl}
          e_n & = & B_{1n}(z)\varepsilon_1 + \Upsilon B_{2n}(z)\varepsilon_2, \\
          \tilde{e}_n & = & B_{1n}^{*}(z)\varepsilon_1 + \Upsilon B_{2n}^{*}(z)\varepsilon_2, \\
          {/\kern -0.50em {\tilde{e}} }_n & = & \gamma^0 {/\kern -0.50em e}_n^{\dagger} \gamma^0,
        \end{array}
        \right.
\end{equation}
and
\begin{equation}
        \label{eq9}
        \left\{
        \begin{array}{rcl}
          u & = & \frac{( k^{'}\cdot k )^2 }{4(q_f \cdot k )({\bar{q}}_i \cdot k)}, \\
          s & = & 2( k^{'} \cdot k ),      \\
          m_{*}^2 & = & m_0^2 + e^2a^2 \Lambda.
        \end{array}
        \right.
\end{equation}
Then we can easily get $ k \cdot e_n = 0 $, since $ k \cdot
\varepsilon_1 = 0 $ and $ k \cdot \varepsilon_2 = 0 $.

We should deal with $ \overline{M_{fi,n}M_{fi,m}^{*}} $ and $
\overline{M_{fi,m}M_{fi,n}^{*}} $ together, based on that the
final result of $\overline{{\left\vert S_{fi} \right\vert}^2}$ should be
real, thus we can define a convenient quantity $M_{fi;mn}^2 = \frac{1}{2}\left( \overline{M_{fi,n}M_{fi,m}^{*}} + \overline{M_{fi,m}M_{fi,n}^{*}} \right)$. After a tedious and careful calculation, we finally obtain
\begin{equation}
        \label{eq10}
        \begin{array}{rcl}
          M_{fi;mn}^2 & = & \frac{1}{16m_0^2} \left\{ \sum\limits_{\lambda} Tr\left[ (/\kern-0.40em p_f + m_0)M_n(/\kern-0.40em \bar{p}_i - m_0) \gamma^0 M_m^{\dagger} \gamma^0 \right] \right. \\
          &    &  \left. + \sum\limits_{\lambda} Tr\left[ (/\kern-0.40em p_f + m_0)M_m(/\kern-0.40em \bar{p}_i - m_0) \gamma^0 M_n^{\dagger} \gamma^0 \right] \right\} \\
          & = & E_{mn,0} + E_{mn,1} + E_{mn,2} + E_{mn,3},
        \end{array}
\end{equation}
where $E_{mn,0}$, $E_{mn,1}$, $E_{mn,2}$ and $E_{mn,3}$ are defined as
        $$
        \begin{array}{rcl}
        E_{mn,0} & = & \cos{[(n-m)\phi_0]} J_n(z)J_m(z) \\
                 &   & + \eta^2 \cos{[(n-m)\phi_0]} \left[  J_n(z)J_m(z) - \frac{J_{m+1}(z)J_{n+1}(z)}{2} - \frac{J_{m-1}(z)J_{n-1}(z)}{2} \right] (1-2u),
        \end{array}
        $$
        $$
        E_{mn,1} = \frac{s}{4m_0^2} (2N-m-n) \cos{[(n-m)\phi_0]} J_n(z)J_m(z),
        $$
        $$
        \begin{array}{rcl}
        E_{mn,2} & = & (1-\frac{1}{\Lambda}) \eta^2 \cos{[(n-m)\phi_0]} \cos{(2\phi_0)}J_n(z)J_m(z) \times \\
                 &   & \left\{ \left[1-2u \right] \left[\frac{mn}{z^2} -\frac{d}{dz}\ln{\left|J_m(z) \right|} \frac{d}{dz}\ln{\left| J_n(z) \right| }\right]
                 - \left[ \frac{n^2+m^2}{z^2} - 1 - \frac{1}{z} \frac{d}{dz}\ln{\left| J_n(z)J_m(z)\right|} \right] \right\},
        \end{array}
        $$
and
        $$
        \begin{array}{rcl}
        E_{mn,3} & = & -(1-\frac{1}{\Lambda}) \eta^2 \sin{[(n-m)\phi_0]} \sin{(2\phi_0)}J_n(z)J_m(z) \times \\
                 &   & \left\{ (n-m) \left[\frac{d}{zdz}\ln{\left| J_n(z)J_m(z)  \right|} - \frac{1}{z^2} \right] - \frac{2u}{z} \frac{d}{dz}\ln{\frac{{\left| J_m(z) \right|}^n}{{\left| J_n(z) \right|}^m}} \right\}.
        \end{array}
        $$
It is necessary to use the basic Bessel relations and the
equation, $p_f \cdot \bar{p}_i = \frac{1}{2}Ns - 2 \eta^2 m_0^2 u
- m_0^2$, to obtain Eq. (\ref{eq10}).

Finally, we get the differential cross section,
\begin{equation}
        \label{eq11}
        \begin{array}{rcl}
        d\bar{\sigma} & = & \frac{1}{{|\vec{J}_{in}|} \rho_{\omega} } \frac{\overline{{\left\vert S_{fi} \right\vert}^2}}{VT} V\frac{d^3q_f}{(2\pi)^3} V\frac{d^3\bar{q}_i}{(2\pi)^3}  \\
               &  = & \frac{e^2}{4\pi} \frac{m_0^2}{q_f^0 \bar{q}_i^0 \omega^{'} } \frac{1}{\rho_{\omega}} d^3q_f d^3\bar{q}_i \times \\
               &    &     \sum\limits_{m_1, n_1=-\infty}^{\infty} \sum\limits_{m_2, n_2=-\infty}^{\infty} \delta\left( q_{f;\mu} + \bar{q}_{i; \mu} - k^{'}_{\mu} -Nk_{\mu} \right) \times \delta_{n_1+2m_1,N} \delta_{n_2+2m_2,N} \left[  M_{fi;n_1n_2}^2 J_{m_1}(y) J_{m_2}(y) \right],
        \end{array}
\end{equation}
where $ \rho_{\omega} $ is defined as $\frac{\Lambda a^2
\omega}{4\pi}$, $\delta_{a,b} = 1$ when $a=b$ and $0$ when
$a\neq b$.

To derive the total cross section $\bar{\sigma}$ of pair
production, we will first only consider the interaction between
the gamma photon and photons taken from the laser field, in which $
q_f + \bar{q}_i = k^{'} + Nk $ is satisfied. To deal with
Eq. (\ref{eq11}), we work in the center coordinate of the pair
created. Then the following quantities could be represented as
\begin{equation}
        \label{eq12}
        \left\{
            \begin{array}{rcl}
                q_f^{\mu} & = & (q_f^0, \vec{q}_f) = (q^0, \vec{q}), \\
                \bar{q}_i^{\mu} & = & ( q^0, \vec{\bar{q}}_i ) = (q^0, -\vec{q}),  \\
                q^2 & = & m_0^2+\Lambda^2e^2a^2 = q_0^2-|\vec{q}|^2, \\
                \omega^{'} & = &  N\omega, q_0 = \omega^{'}.
            \end{array}
            \right.
\end{equation}
and
\begin{equation}
        \label{eq13}
        \left\{
            \begin{array}{rcl}
                s & = & 4 \omega^{'} \omega = 4N\omega^2, \\
                q_f \cdot k & = & \omega^{'} \omega - |\vec{q}| \omega \cos{\theta}, \\
                u & = & \frac{Ns}{4\left[ q_0^2 - |\vec{q}|^2\cos^2{\theta}  \right]}, \\
                \cos{\theta} & = & \pm \frac{q_0}{|\vec{q}|} \sqrt{1-\frac{1}{u}}.
            \end{array}
            \right.
\end{equation}
Here, we first write a useful integral, $\int\int \, d^3q_f
d^3\bar{q}_i \frac{1}{q_f^0 \bar{q}_i^0}\delta\left( q_{f;\mu} +
\bar{q}_{i; \mu} - k^{'}_{\mu} -Nk_{\mu} \right)$, which in fact
could be treated as an integral operator, into a more compact
form. With the help of Eq. (\ref{eq2}), Eq. (\ref{eq12}),
Eq. (\ref{eq13}) and the property of Dirac function $\delta$, we
have
        $$
        \begin{array}{l}
            \int\int\, d^3q_f d^3 \bar{q}_i \frac{1}{q_f^0 \bar{q}_i^0} \delta \left( q_{f;\mu} + \bar{q}_{i; \mu} - k^{'}_{\mu} -Nk_{\mu} \right)
            = \int\, d^3q \frac{1}{q_0^2} \delta \left( 2q_0 - \omega^{'} - N\omega \right) \\
            = \int\, d{\varphi} \int\, d{\cos{\theta}} \int\,d\left|\vec{q}\right| \frac{ {\left| \vec{q} \right| }^2}{q_0^2} \delta \left( 2q_0 - \omega^{'} - N\omega \right) = \int\, d{\varphi} \int\, d{\cos{\theta}}  \frac{ {\left| \vec{q} \right| } }{2q_0} \\
            =\frac{1}{2} \int\,d\varphi \int\, du \frac{1}{u\sqrt{u(u-1)}}.
        \end{array}
        $$
Then the total cross section $\bar{\sigma}_N$ of the pair production,
in the interaction between one high energy gamma and $N$ photons
of the laser field, can be presented in a compact form,
\begin{equation}
        \label{eq14}
        \begin{array}{rcl}
             \bar{\sigma}_N & = & \frac{2\alpha^2}{\vartheta m_0^2 \eta^2} \int^{2\pi}_{0} \,d\varphi \int^{u_N}_1 \,  du \frac{1}{u\sqrt{u(u-1)}} \times \\
            & & \sum\limits_{m_1, n_1=-\infty}^{\infty} \sum\limits_{m_2, n_2=-\infty}^{\infty}\delta_{n_1+2m_1, N } \delta_{n_2+2m_2, N } \left[  M_{fi;n_1n_2}^2 J_{m_1}(y) J_{m_2}(y) \right],
        \end{array}
\end{equation}
where $\alpha$ is the fine-structure constant, $ \vartheta =
\frac{s}{m_0^2} $ and $ u_N = \frac{N\vartheta}{4(1+\eta^2)} $ with $N>n_0$ because the minimum photon number $n_0$
is need to overcome the energy gap to create the $e^+ e^-$ pair.
The total cross section $\bar{\sigma}$ is
$\sum\limits^{+\infty}_{N=n_0} \bar{\sigma}_N$. It is easy to get
from Eq. (\ref{eq12}) and Eq. (\ref{eq13}) the relation, $
\frac{Ns}{4\left[ q_0^2+|\vec{q}|^2 \right]} \le u \le
\frac{Ns}{4\left[ q_0^2 - |\vec{q}|^2 \right]}  $, i.e. $ u \in
[1, u_N] $.

It is noted that Eq. (\ref{eq14}) is determined by only three
parameters, $\Upsilon$, $\eta$ and $ \vartheta $. The remaining
problem is how to handle $ \phi_0 $, $y$ and $z$. It is possible
to get their final expressions from the relations in
Eq. (\ref{eq2}), Eq. (\ref{eq9}), Eq. (\ref{eq12}) and
Eq. (\ref{eq13}). After a careful calculation with these relations,
we finally obtain
\begin{equation}
        \label{eq15}
        \left\{
        \begin{array}{rcl}
            y & = & 2(1-\frac{1}{\Lambda}) \frac{\eta^2 u}{\vartheta}, \\
            z & = & \frac{8\eta}{\sqrt{\Lambda} \vartheta} \sqrt{ (1+\eta^2) u \left[ u_N-u \right] + 2(\Lambda - 1)\left[ \frac{N\vartheta}{4} - (1+\eta^2)u \right] u \sin^2\varphi  },  \\
            \cos\phi_0 & = & \frac{8u\eta}{z\vartheta\sqrt{\Lambda}} \sqrt{\frac{N\vartheta}{4u} - (1+\eta^2) } \cos\varphi,  \\
            \sin\phi_0 & = & \frac{8\Upsilon u \eta}{z\vartheta\sqrt{\Lambda}} \sqrt{\frac{N\vartheta}{4u} - (1+\eta^2) } \sin\varphi.
        \end{array}
        \right.
\end{equation}
As a check we get easily the result for the circular polarization,
        $$
        \begin{array}{rcl}
            \bar{\sigma} & = & \sum\limits^{+\infty}_{N=1} \bar{\sigma}_N  \\
            & = &  \frac{4\pi \alpha^2}{\vartheta m_0^2 \eta^2} \int\limits^{u_N}_1 \,  du \frac{1}{u\sqrt{u(u-1)}} \times \\
            & & \sum\limits^{\infty}_{N \ge n_0}  \left\{ J_N^2(z)+\eta^2\left[ J_N^2(z) - \frac{J_{N+1}^2(z) + J_{N-1}^2(z) }{2} \right] (1-2u)  \right\},
        \end{array}
        $$
where $n_0$ is the minimum integer among possible values for $N$ that satisfy the
condition $u_N>1.0$. It is also the minimum number to overcome the energy gap for pair production mentioned above.
Obviously our results for this special case is the same as that in Ref. \cite{GrRe}.

\section{NUMERICAL RESULT}

Based on the theoretical formula obtained in last section we can
calculate numerically the cross section for the pair production
for different parameters. The C code of Bessel function in Ref.
\cite{HTVF} is used in our program.

With Eq. (\ref{eq14}), the corresponding cross section in MKSA units should be
\begin{equation}
        \label{eq16}
        \begin{array}{rcl}
             \bar{\sigma}_N & = & \left( \frac{\hbar}{m_0 c} \right)^2 \frac{2\alpha^2}{\vartheta \eta^2} \int\limits^{2\pi}_{0} \,d\varphi \int\limits^{u_N}_1 \,  du \frac{1}{u\sqrt{u(u-1)}} \times \\
            & & \sum\limits_{m_1, n_1=-\infty}^{\infty} \sum\limits_{m_2, n_2=-\infty}^{\infty}\delta_{n_1+2m_1, N } \delta_{n_2+2m_2, N } \left[  M_{fi;n_1n_2}^2 J_{m_1}(y) J_{m_2}(y) \right],
        \end{array}
\end{equation}
where $N \geq n_0$.

Our numerical simulations are based on Eq. (\ref{eq16}), and the
numerical results have been scaled by $\frac{2\alpha^2
\hbar^2}{m_0^2c^2}$, then the total cross section $\bar{\sigma}$
is $\sum\limits^{+\infty}_{N=1} \bar{\sigma}_N$. The main
numerical results will be demonstrated graphically in the
following subsections. It is noted that we have also reproduced
the numerical result of Ref. \cite{GrRe} for the special case, i.e.
the circular polarization.

\subsection{Pair production cross section dependence on laser field polarization and intensity}

\begin{figure}[htbp]\suppressfloats
\includegraphics[width=12cm]{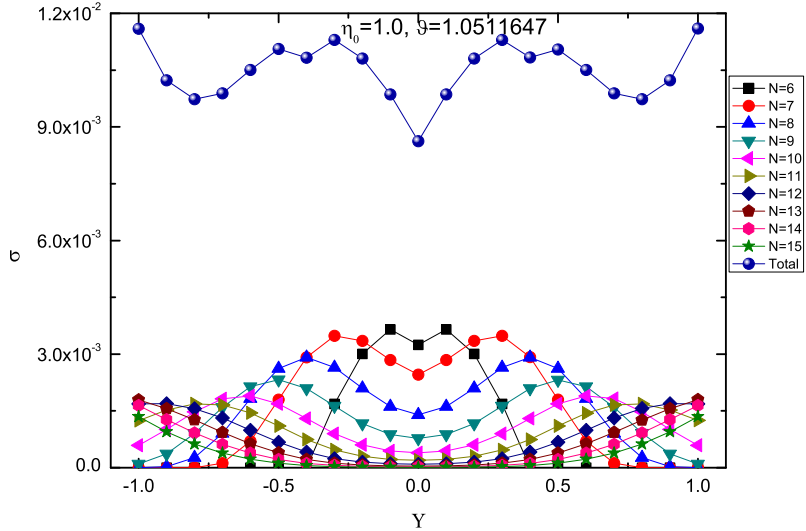}
\caption{\label{fig1} (Color online)  The pair cross section $\bar{\sigma}_N$ for
different number of photons dressed, i.e, from 6 to 30 and $
\bar{\sigma}$, the summation of them, are plotted. Here, for example,
$N=6$, represents the number of photons dressed from the laser
field in the multi-photons process. $\eta_0$ and $ \vartheta
$ are the parameters fixed in the numerical simulation, while
$\Upsilon$ is controlled. We only plot $\bar{\sigma}_N$ for $ N
\in \{6, 7,\ldots ,15\}$, and $\bar{\sigma} \approx
\sum^{30}_{N=6} \bar{\sigma}_N$, because $\bar{\sigma}_N = 0$ when
$N=\{ 1, \ldots, 5\}$. All other figures will have these
adoptions unless otherwise specified.}
\end{figure}

We consider first the laser field applied in the SLAC experiment
and give an insight about the experiments with $\vartheta =
1.0511647$ and $\eta_0 = ea/m_0 = 1.0 $. Note that it is different
from $\eta = \eta_0 \sqrt{\Lambda}$ in our general case since the
laser field polarization affects $\Lambda$. The relation between
the cross sections of pair production and the parameter $\Upsilon$
is displayed in Fig. \ref{fig1}.

\begin{figure}[htbp]\suppressfloats
\includegraphics[width=12cm]{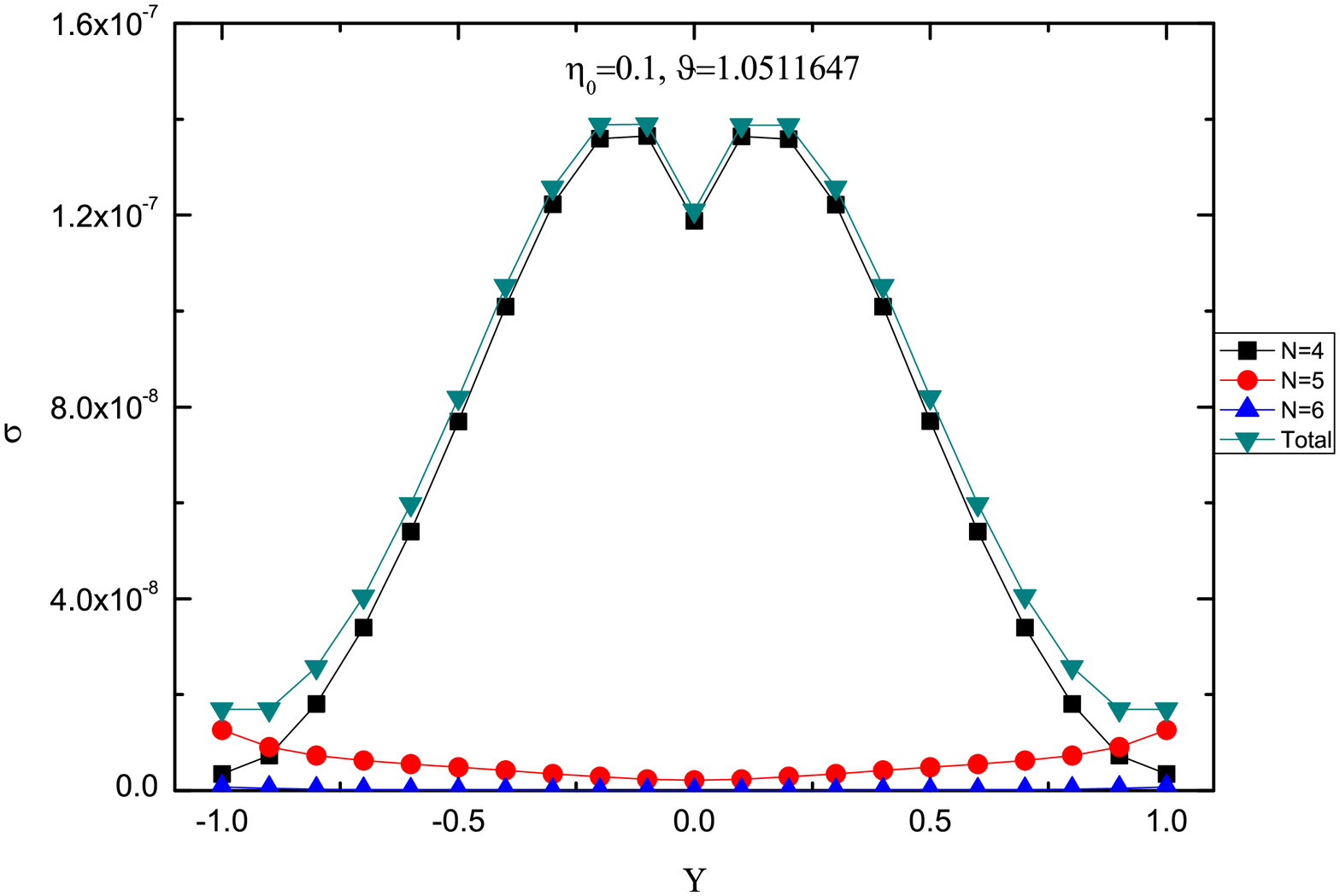}
\caption{\label{fig2} (Color online)  Pair production
$\bar{\sigma}_N$ and $\bar{\sigma}$ when $\eta_0 = 0.1$ and
$\vartheta = 1.0511647$. Here $\bar{\sigma} \approx
\sum^{10}_{N=4} \bar{\sigma}_N$ and $\bar{\sigma}_N$ for $ N \in
\{7,\ldots ,10\}$ have been omitted for their insignificance.}
\end{figure}

Surprisingly we can clearly see that the optimal polarization
parameter $\Upsilon$ is $1.0$ or $-1.0$ rather than $0$ as thought by some
people. On the other hand, the total production cross sections
$\bar{\sigma}$ at $\Upsilon = 0.3$, $0.4$ and $0.5$ are also very
close to the optimal one. Furthermore, in the case of linear
polarization the total pair production cross section is the
smallest. It can be easily obtained that $\bar{\sigma}(\Upsilon
=1) / \bar{\sigma}(\Upsilon =0) \approx 1.34 $, and
$\bar{\sigma}(\Upsilon =0.3) / \bar{\sigma}(\Upsilon =0)  \approx
1.31$. These ratios indicate that the parameter $\Upsilon$ plays
a very important role in directing the experiments to get more
obvious and easy-detected results. We can also see that the
numerical result is symmetric with respect to $\Upsilon = 0$ as is
expected by the theoretical requirement in Eq. (\ref{eq16}).

\begin{figure}[htbp]\suppressfloats
\includegraphics[width=12cm]{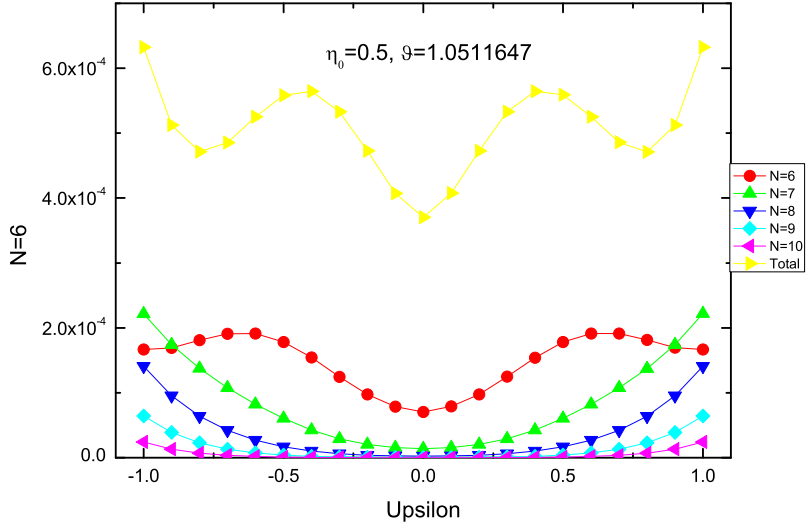}
\caption{\label{fig3} (Color online)  Pair production
$\bar{\sigma}_N$ and $\bar{\sigma}$ when $\eta_0 = 0.5$ and
$\vartheta = 1.0511647$. Here $\bar{\sigma} \approx
\sum^{14}_{N=5} \bar{\sigma}_N$ and $\bar{\sigma}_N$ for $ N \in
\{5,11,\ldots ,14\}$ have been omitted for their insignificance.}
\end{figure}

\begin{figure}[htbp]\suppressfloats
\includegraphics[width=12cm]{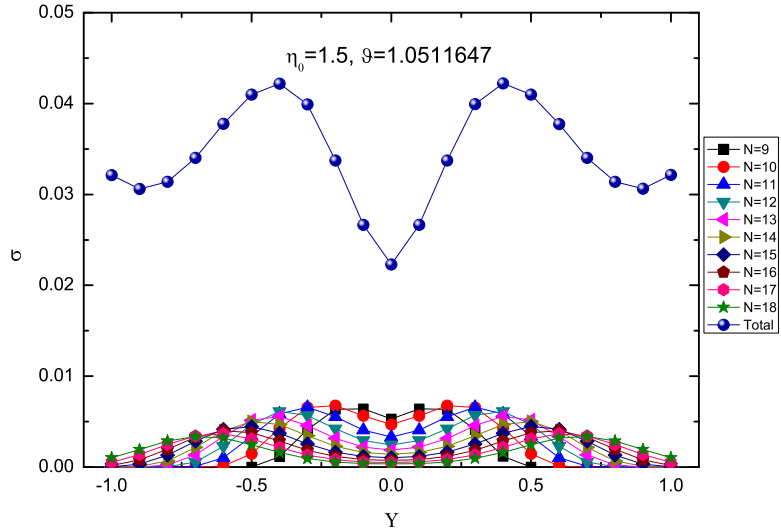}
\caption{\label{fig4} (Color online)  Pair production
$\bar{\sigma}_N$ and $\bar{\sigma}$ when $\eta_0 = 1.5$ and
$\vartheta = 1.0511647$. Here $\bar{\sigma} \approx
\sum^{40}_{N=9} \bar{\sigma}_N$ and $\bar{\sigma}_N$ for $ N \in
\{19,\ldots ,40\}$ have been omitted, but the contribution from
them is not insignificant.}
\end{figure}

\begin{figure}[htbp]\suppressfloats
\includegraphics[width=12cm]{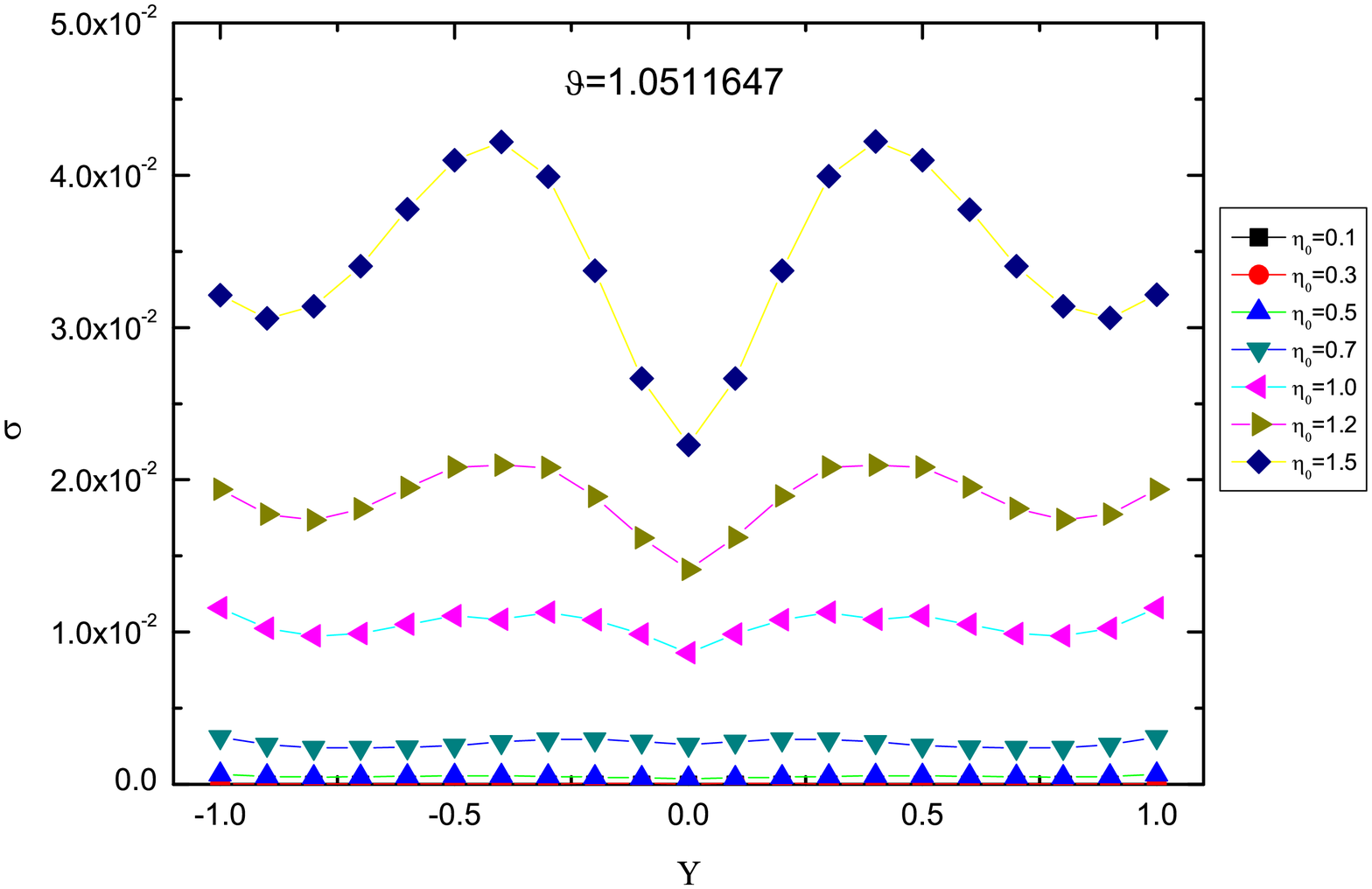}
\caption{\label{fig5} (Color online)   All the total cross
sections $\bar{\sigma}$ for different $\eta_0$ when $\vartheta =
1.0511647$ are plotted in one figure to unearth more
information.}
\end{figure}

Then we will unearth how the cross sections, $\bar{\sigma}_N$ and
$\bar{\sigma}$, are influenced by $\eta_0$, the intensity of the
polarized laser field. Usually the parameter $\vartheta$ is kept
as a constant in experiments when $\eta_0$ is changed. Now we will
calculate the pair production cross sections for different
$\eta_0$. Here the strong constraint, $u_N =
\frac{N\vartheta}{4(1+\eta^2)} > 1 $, should be taken into account
seriously, otherwise the integral of Eq. (\ref{eq16}) would be
zero. Given $\vartheta=1.0511647$ some numerical results with
$\eta_0 = 0.1$, $0.5$ and $1.5$ are plotted in Fig. \ref{fig2},
Fig. \ref{fig3} and Fig. \ref{fig4}, respectively. We also give the
total cross sections $\bar{\sigma}$ for a series of different
field intensities in Fig. \ref{fig5} in order to see the effect of
$\eta_0$.

\begin{table}[htbp]
        \caption{\label{table1} The relation between the optimal $\Upsilon$ which makes $\bar{\sigma}_N$ maximum and $N$ when $\eta_0 = 1.0$ and $\vartheta = 1.0511647$. "0.9/1.0" means that the cross section $\bar{\sigma}_N$ when $\Upsilon = 0.9$ and $\Upsilon = 1.0$ cannot be distinguished in our numerical simulation. Similar in Tables \ref{table2} and \ref{table3}.}
        \begin{ruledtabular}
        \begin{tabular}{ccccccccccc}
            $N$ & 6 & 7 & 8 & 9 & 10 & 11 & 12 & 13 & 14 & 15 \\
            $|\Upsilon|$ & 0.1 & 0.3 & 0.4 & 0.5 & 0.6 & 0.7 & 0.9/1.0 & 1.0 & 1.0 & 1.0
        \end{tabular}
        \end{ruledtabular}
    \end{table}

As an example, in Table \ref{table1} we list the values of $|\Upsilon|$
corresponding to the optimal pair production cross section
$\bar{\sigma}_N$ for different $N$ when $\vartheta=1.0511647$ and
$\eta_0 = 1.0$. Actually, we have studied all the data we have got
from numerical calculations, and find a fact that there exists a
tendency: the larger $N$, the greater $|\Upsilon|$. This means that the
polarized field becomes closer to the circular one for the optimal
pair production $\bar{\sigma}_N$ as $N$ increases. This behavior is more obvious
for smaller $\eta_0$ where $N$ is larger. Therefore, a dominant
contribution to $\bar{\sigma}_N$ arises from $|\Upsilon|= 1$
when $N$ is large. However, this conclusion for total pair
production is not valid in general because
the $N$-order cross section $\bar{\sigma}_N$ attributed to the total cross section
decreases as $N$ increases. This point is also seen in Fig. \ref{fig5}, for
example, where the optimal total pair production for $\eta_0=1.5$ is not
at $|\Upsilon|= 1$ but at $0.4$. It should be emphasized that for the total
cross section there is a trade-off between $\eta_0$ and $\Upsilon$
which makes $\bar{\sigma}$ optimal.

\begin{table}[htbp]
        \caption{\label{table2} The relation between the optimal $\Upsilon$ which makes $\bar{\sigma}$ maximum and $\eta_0$ when $\vartheta = 1.0511647$.}
        \begin{ruledtabular}
        \begin{tabular}{ccccccccccc}
            $\eta_0      $ & 0.1 & 0.3 & 0.5 & 0.7 & 1.0 & 1.2         & 1.5 \\
            $|\Upsilon|    $ & 0.1 & 0.1 & 1.0 & 1.0 & 1.0 & 0.3/0.4/0.5 & 0.4
        \end{tabular}
        \end{ruledtabular}
    \end{table}

    \begin{table}[htbp]
        \caption{\label{table3} The relation between the optimal $\Upsilon$ which makes $\bar{\sigma}$ maximum and $\vartheta$ when $\eta_0 = 1.0$.}
        \begin{ruledtabular}
        \begin{tabular}{ccccccccccc}
            $\vartheta   $ & 1.0511647 & 3.0     & 6.0 & 10.0 & 15.0 & 21.0 & 28.0 & 36.0    & 45.0    & 55.0 \\
            $|\Upsilon|    $ & 1.0       & 0.5/0.6 & 0.6 & 0.3  & 0.3  & 0.3  & 0.3  & 0.2/0.3 & 0.2/0.3 & 0.2/0.3
        \end{tabular}
        \end{ruledtabular}
    \end{table}

On one hand, for smaller $\eta_0$ the main contribution to
total $\bar{\sigma}$ comes from $\bar{\sigma}_{n_0}$ since
other terms $(\bar{\sigma}_{N>n_0})$ decrease rapidly, see
Fig. \ref{fig2}. In this situation the optimal polarization
parameter approaches the linear one, see also Table \ref{table2} in the cases of
$\eta_0=0.1$ and $0.3$. On the other hand, as $\eta_0$ increases
to $\sim 1.0$, the field polarization parameter of optimal
total pair production cross section becomes the circular one, $|\Upsilon|= 1$.
The same results are shown in Table \ref{table1} for very large $N$, because
in these cases the terms of $\bar{\sigma}_{N>n_0}$ decrease more slowly so that the whole
contribution from these large $N$ terms exceeds that from those
small $N$ terms. This point can also be seen clearly in
Fig. \ref{fig1} and Fig. \ref{fig3}. However, as $\eta_0$ increases
further a typical nonlinear feature arises between $|\Upsilon|$
and $\eta_0$ for the optimal pair creation. From Fig.  \ref{fig4}
and Table \ref{table2} the trade-off between $|\Upsilon|$ and
$\eta_0$ contribution to total cross section compels the optimal
polarization to locate at about $|\Upsilon|= 0.4$, which is between
linear and circular polarizations for our studied field parameters.

In a word, although the pair production cross section increases
monotonically with laser field intensity, its optimal value
depends strongly on laser field polarization, see Fig. \ref{fig5}.
The former can be understood by the perturbation concept but the
later can be recognized as the typical nonlinear interaction
between fields and vacuum in QED, i.e. the laser dressed
electrons/positrons colliding with a high energy photon.

\subsection{Pair production cross section dependence on $\vartheta$}
\begin{figure}[htbp]\suppressfloats
\includegraphics[width=12cm]{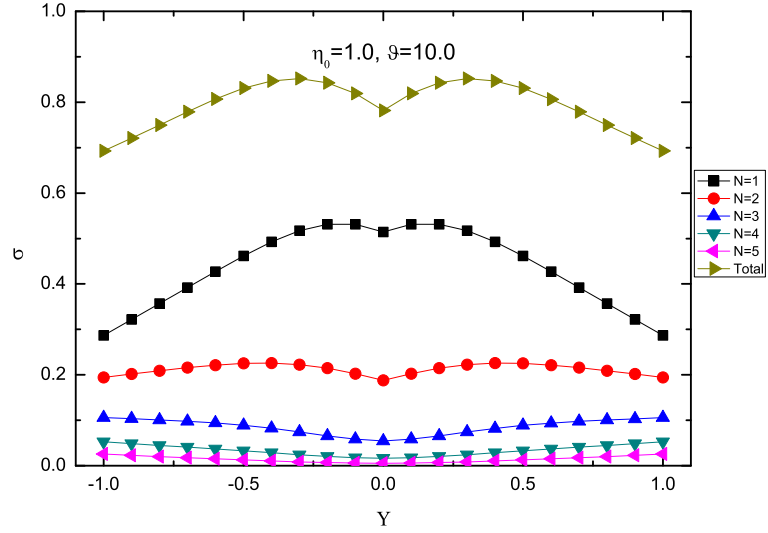}
\caption{\label{fig6} (Color online)  Pair production
$\bar{\sigma}_N$ and $\bar{\sigma}$ when $\eta_0 = 1.0$ and
$\vartheta = 10.0$. Here $\bar{\sigma} \approx \sum^{10}_{N=1}
\bar{\sigma}_N$ and $\bar{\sigma}_N$ for $ N \in \{6,\ldots ,10\}$
have been omitted for their insignificance.}
\end{figure}

\begin{figure}[htbp]\suppressfloats
\includegraphics[width=12cm]{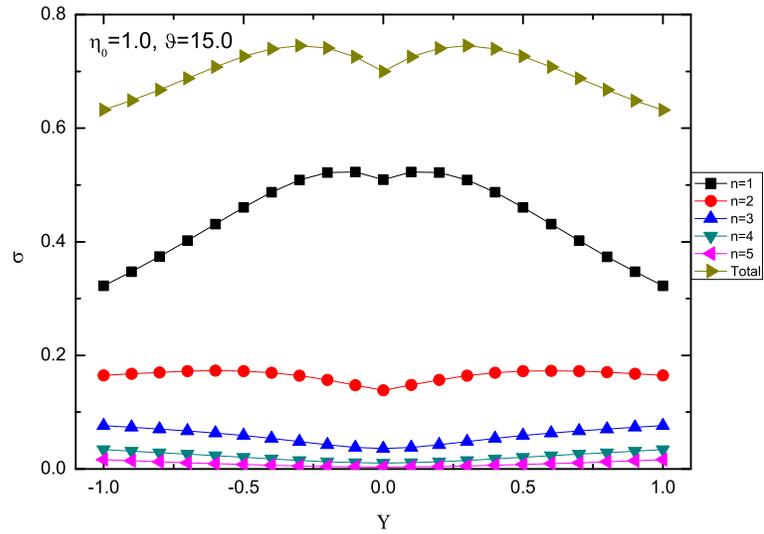}
\caption{\label{fig7} (Color online)  Pair production
$\bar{\sigma}_N$ and $\bar{\sigma}$ when $\eta_0 = 1.0$ and
$\vartheta = 15.0$. Here $\bar{\sigma} \approx \sum^{10}_{N=1}
\bar{\sigma}_N$ and $\bar{\sigma}_N$ for $ N \in \{6,\ldots ,10\}$
have been omitted for their insignificance.}
\end{figure}

\begin{figure}[htbp]\suppressfloats
\includegraphics[width=12cm]{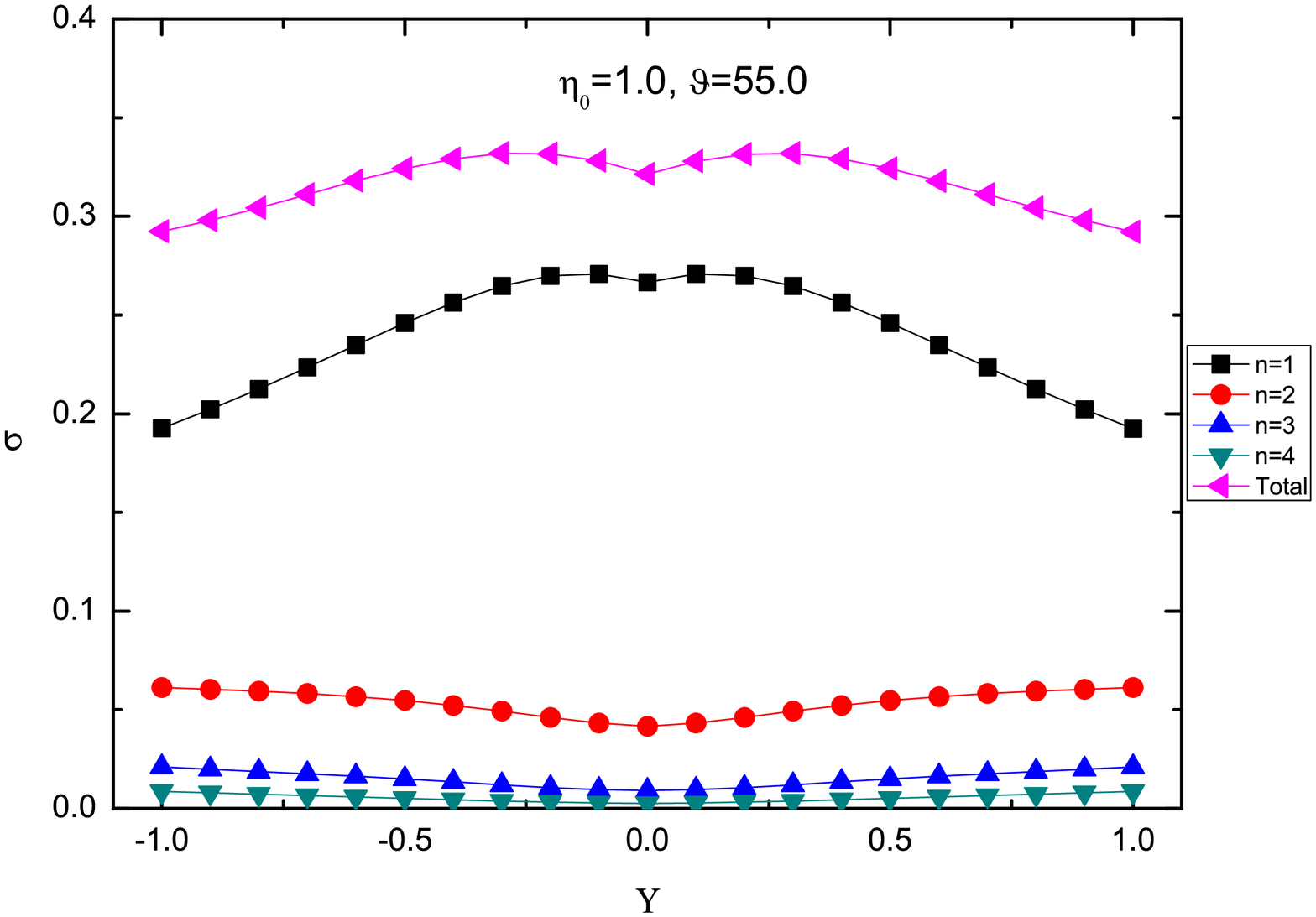}
\caption{\label{fig8} (Color online)  Pair production
$\bar{\sigma}_N$ and $\bar{\sigma}$ when $\eta_0 = 1.0$ and
$\vartheta = 55.0$.  Here $\bar{\sigma} \approx \sum^{10}_{N=1}
\bar{\sigma}_N$ and $\bar{\sigma}_N$ for $ N \in \{5,\ldots ,10\}$
have been omitted for their insignificance.}
\end{figure}

\begin{figure}[htbp]\suppressfloats
\includegraphics[width=12cm]{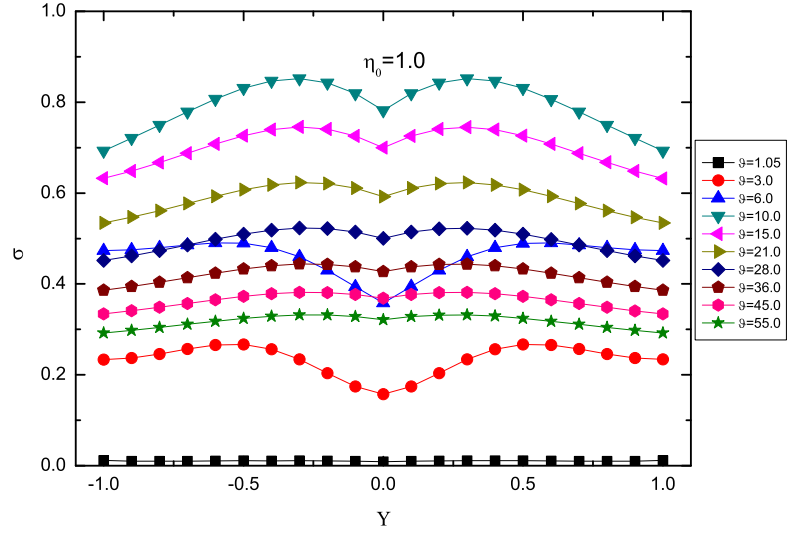}
\caption{\label{fig9} (Color online)   All the total cross
sections $\bar{\sigma}$ in different $\vartheta$ when $\eta_0 =
1.0$ are plotted in one figure to unearth more information.
Here $\vartheta = 1.0511647$ is shortened as $\vartheta = 1.05$ in the figure.}
\end{figure}

\begin{figure}[htbp]\suppressfloats
\includegraphics[width=12cm]{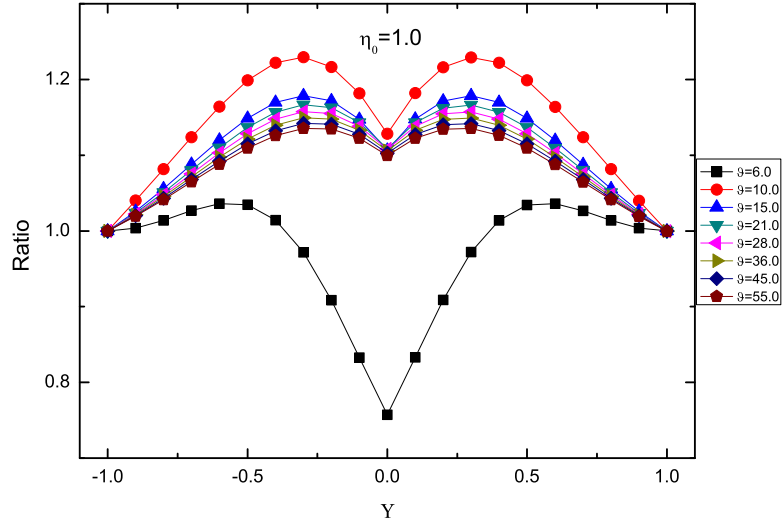}
\caption{\label{fig10} (Color online)  This figure is
corresponding to Fig. \ref{fig9} and here "Ratio" is the ratio
of $\bar{\sigma}_{\Upsilon}$ and the corresponding $\bar{\sigma}$
of the circular in the same parameters $\vartheta$ and $\eta_0$.
We have omitted the cases when $\vartheta = 1.0511647, 3.0$.}
\end{figure}

Besides two important parameters, $\Upsilon$ and $\eta_0$, the pair creation is also closely
related to the parameter $\vartheta$. The problem, in fact, is
how to tune the parameter $\vartheta$, to improve the pair
production, given that $ u_N = \frac{N\vartheta}{4(1+\eta^2)}>1 $.
The idea is conformed with the physical picture as
$\sqrt{\vartheta} = \sqrt{2(k^{'}\cdot k)}/m_0$, which means
the collision energy of two photons. Usually one believes that
the larger $\vartheta$ is, the more likely the process of pair
generation occurs. But things may not be so simple. We will find that the pair production will surely be
greatly enhanced, in the first order, then begin to decrease in
quantity when $\vartheta$ is bigger than a critical $\vartheta_c$,
which again exhibits the nonlinear characteristic of QED process. Some
values of $\vartheta$ are selected for fixed $\eta_0$, which are
3.0, 6.0, 10.0, 15.0, 21.0, 28.0, 36.0, 45.0 and 55.0 and the
numerical results with $\vartheta = 10.0, 15.0, 55.0$ are
demonstrated graphically from Fig. \ref{fig6} to Fig. \ref{fig8}. We
also give all the total cross sections $\bar{\sigma}$  for
different $\vartheta$ in Fig. \ref{fig9} and Fig.  \ref{fig10} in order to
make convenient comparisons with each other.

Now we will see how $\vartheta$ influences the pair
creation. Firstly, from Fig. \ref{fig9}, we can see that the pair
production is greatly enhanced with the increase of $\vartheta$
for fixed $\eta_0$, same behavior as shown in
Ref. \cite{GrRe} for the circularly polarized field. But it is not
monotonically increasing with $\vartheta$ since the total cross
section $\bar{\sigma}$ starts to decrease when $\vartheta \ge
15.0$ in our numerical results. The physical mechanism of
this phenomenon seems difficult to grasp, however, a closer
inspection of the integral of Eq. (\ref{eq16}) shows that there exist two
mathematic reasons for this counterintuitive phenomenon, i.e. the influence of
Bessel functions and the inverse relation between the total cross
section $\bar{\sigma}$ and $\vartheta$ in the coefficient
$\frac{1}{\vartheta\eta^2}$. Certainly whether this phenomenon is kept in multiphoton perturbation regime,
in which the higher order terms are included, or even in nonperturbation regime, is still an open problem.
This difficult problem may be overcome in future possible theoretical calculations as well as more experiments.

$\bar{\sigma}$ is a summation of different $\bar{\sigma}_N$
when $N$ is subject to $u_N > 1.0$. Let us remember a fact that if the
main contribution comes from the first term with least
number $N=n_0$ then the optimal pair production is prone to the linearly polarized
field. This is more obvious for larger $\vartheta$ as shown in Table \ref{table3}. But there are
some exceptions for smaller $\vartheta$, for example, $|\Upsilon|
= 1$ corresponds to parameter $\vartheta \approx 1.05$ in SLAC experiment
case. More interesting, in our first order perturbation theory, when
$\vartheta$ is large the pair production seems insensitive to the
polarization parameter $\Upsilon$. From our numerical results we can
see clearly, in Fig. \ref{fig1}, Fig. \ref{fig9} and
Fig. \ref{fig10}, that on one hand there is relatively large fluctuation of the cross
section with respect to the polarization parameter for $\vartheta \in [1.05, 21.0]$
and on the other hand there is indeed relatively small change of the cross section
with respect to the polarization parameter $\Upsilon$ when $\vartheta \ge 28.0$.
Although the main reason is unclear physically, the conclusion
from numerical results is still very meaningful and useful.
It provides an intuitive conclusion that when
the collision energy of two photons is large enough usually one can
directly choose the linearly or circularly polarized laser field,
which is relatively easier to get in laboratories, for the pair production. With
such special choices of polarization the cross section of the pair production does not
differ from the optimal one greatly (less than 15\% when $\vartheta \ge 36.0$, as shown in Fig. \ref{fig10}).

\section{Discussions and conclusions}
We have obtained the rigorous formula in the first order
perturbation theory and performed thorough numerical
computations for the pair production. Our numerical results for the
parameters of the SLAC experiment have clearly manifested that the linear
laser field is not the optimal one, but rather the circular one,
although the partial cross section $\bar{\sigma}_6$ in the case of
linear polarization is larger than that in the case of circular one.
Surprisingly, neither linear nor circular but the elliptically polarized field
corresponds to the optimal pair creation cross section in some
cases shown in Table \ref{table2} and Table \ref{table3}.

Our numerical results have also demonstrated that the leading
contribution in the pair production comes from the first term subject to
$u_N = \frac{N\vartheta}{4(1+\eta^2)} > 1.0$, when the polarization is nearly linear;
moreover, things begin to change as larger $N$
is considered, since $\bar{\sigma}_N$ for larger $N$ is prone to
the circular one. It might be the main reason that causes the balance
between the linearly and the circularly polarized and also the
reason of many nonlinear dependence of pair production on other system parameters.
From our numerical results, we also see that when $\eta_0$ is large the contributions from
$\bar{\sigma}_N$ with large $N$ are very important. While
$\vartheta$ is large the contribution from $\bar{\sigma}_N$ with
large $N$ are not so important. For example, given $\eta_0 = 1.0$, when
we calculate the total cross section for $\vartheta = 1.0511647$,
at least $\sum^{30}_{N=6} \bar{\sigma}_N$ is used; however, for
$\vartheta \ge 6.0$, summation by $\sum^{10}_{N=1}
\bar{\sigma}_N$ as an approximation of $\bar{\sigma}$ is already good
enough. By the way for $\eta_0 = 1.5$ and $\vartheta = 1.0511647$
a calculation by $\sum^{40}_{N=9} \bar{\sigma}_N$ is needed.

We have exhibited the abundance phenomena in the process of
photon-multiphoton reaction that the laser field polarization plays a key role
in optimal pair production when the other parameters
are fixed, which can be seen from Fig.  \ref{fig1} to Fig.
\ref{fig10} and three tables. Our work may provide some
understanding about some previous experimental phenomena \cite{Burke,Altarelli}
and also direction for future experiments. Although a great effort has been made to
understand the polarization effect on the pair production, there still remain
some open problems to be solved, especially the physical
mechanism behind the complex calculational formulae. We conjecture
that there may be nonlinear resonance or/and interference terms
appeared in different modes of scattering matrix elements for
the transition amplitude which would enhance or reduce the pair
production cross section in different polarization cases. However,
more theoretical and experimental research are still needed in the
future to have a deeper understanding of vacuum decay in
ultrastrong laser field to create electron-positron pairs, which serves as an
important test for nonlinear QED.

\begin{acknowledgments}
This work was supported by the National Natural Science Foundation
of China (NNSFC) under the grant Nos. 11175023, 10975018 and
11175020, and partially by the Fundamental Research Funds for the
Central Universities (FRFCU).
\end{acknowledgments}


\begin{thebibliography}{99}\suppressfloats
\bibitem{StMo}
D. Strickland and G. Mourou, Opt. Commum. {\bf 56}, 219(1985).

\bibitem{MoTaBu}
G. Mourou, T.Tajima and S. Bulanov, Rev. Mod. Phys. {\bf 78}, 309(2006).

\bibitem{Burke}
D. L. Burke \emph{et al.}, Phys. Rev. Lett. {\bf 79}, 1626(1997).

\bibitem{Bamber}
C. Bamber \emph{et al.}, Phys. Rev. D {\bf 60}, 092004 (1999).

\bibitem{ELI}
http://www.extreme-light-infrastructure.eu/

\bibitem{Schwinger}
J. Schwinger, Phys. Rev. {\bf 82}, 664(1951).

\bibitem{Volkov}
D. M. Volkov, Z. Phys. {\bf 94}, 250(1935).

\bibitem{Sauter}
F. Sauter, Z. Phys. {\bf 69}, 742(1931).

\bibitem{KimPa}
S. P. Kim and D. N. Page, Phys. Rev. D {\bf 65}, 105002(2002).

\bibitem{DuSc}
G. V. Dunne and C. Schubert, Phys. Rev. D {\bf 72}, 105004(2005).

\bibitem{GiKl}
H. Gies and K. Klingm\"{u}ller, Phys. Rev. D {\bf 72}, 065001(2005).

\bibitem{DuQHWGi}
G. V. Dunne Q. H. Wang, H. Gies and C. Schubert, Phys. Rev. D {\bf 73}, 065028(2006).

\bibitem{MVG}
C. Ml\"{u}ler, A. B. Voitkiv and N. Gr\"{u}n, Phys. Rev. A {\bf 67}, 063407(2003).

\bibitem{DeM}
C. Deneke and C. M\"{u}ller, Phys. Rev. A {\bf 78}, 033431(2008).

\bibitem{MHaKe}
C. M\"{u}ller, K. Z. Hatsagortsyan and C. H. Keitel, Phys. Rev. A {\bf 78}, 033408(2008).

\bibitem{HuMKe}
H. Y. Hu, C. M\"{u}ller and C. H. Keitel, Phys. Rev. Lett. {\bf 105}, 080401(2010).

\bibitem{Popov}
M. S. Marinov and V. S. Popov, Sov. J. Nucl. Phys. 16, 449(1973).

\bibitem{Bulanov}
S. S. Bulanov, Phys. Rev. E {\bf 69}, 036408(2004).

\bibitem{Reiss}
H. R. Reiss, J. Math. Phys. {\bf 3}, 59(1962).

\bibitem{NiRi}
A. I. Nikishov and V. I. Ritus, Sov. Phys. JETP {\bf 19}, 529(1964).

\bibitem{Altarelli}
M. Altarelli \emph{et al.}, Technical Design Report of the European XFEL, DESY 2006-097(http://www.xfel.net).

\bibitem{GrRe}
W. Greiner and J. Reinhardt, \emph{Quantum Electrodynamics}, p. 222-232 (Springer-Verlag Berlin Heidelberg, Fourth Edition 2009).

\bibitem{BeLiPi}
V. B. Berestetskii, E. M. Lifshitz and L. P. Pitaevskii, \emph{Quantum Electrodynamics} (Pergamon, Oxford, 1982).

\bibitem{HTVF}
W. H. Press, S. A. Teukolsky, W. T. Vetterling and B. P. Flannery,
\emph{Numerical Recipes in C} (Cambridge University Press, Cambridge, 1992).

\end{thebibliography}
\end{document}